\documentstyle[12pt]{article}
\global\parskip 6pt

\normalsize

\def\br{\begin{thebibliography}{abc}}
\def\er{\end{thebibliography}}

\def\be{\begin{equation}}
\def\ee{\end{equation}}
\def\bea{\begin{eqnarray}}
\def\eea{\end{eqnarray}}

\begin{document}
\begin{flushright}
\hfill{HMI 05 06}\\
\hfill{SINP-TNP/05-04}\\
\end{flushright}
\vspace*{1cm}
\thispagestyle{empty}
\centerline{\large\bf Geometric Finiteness and Non-quasinormal Modes
of the BTZ Black Hole }
\bigskip
\begin{center}
Kumar S. Gupta\footnote{Email: kumars.gupta@saha.ac.in, 
Regular Associate, Abdus Salam ICTP, Trieste,
Italy.}\\
\vspace*{.2cm}
{\em Theory Division\\
Saha Institute of Nuclear Physics\\
1/AF Bidhannagar\\
Calcutta - 700064, India}\\
\vspace*{.5cm}
Siddhartha Sen\footnote{Email: sen@maths.tcd.ie}\\
\vspace*{.2cm}
{\em Hamilton Mathematics Institute TCD, Dublin 2, Ireland}\\ 
\vspace*{.1cm}
{\em and}\\
\vspace*{.1cm}
{\em Department of Theoretical Physics}\\
{\em Indian Association for the Cultivation of Science}\\
{\em Calcutta - 700032, India}\\
\end{center}
\vskip.5cm

\begin{abstract}

The BTZ black hole is geometrically finite. This means that its three
dimensional hyperbolic structure as encoded in its metric is in 1-1
correspondence with the Teichmuller space of its boundary which is a two
torus. The equivalence of different Teichmuller parameters related by the
action of the modular group therefore requires the invariance of the
monodromies of the solutions of the wave equation around the inner and outer
horizons in the BTZ background. We show that this invariance condition leads
to the non-quasinormal mode frequencies discussed by Birmingham and Carlip.

\end{abstract}
\vspace*{.3cm}
\begin{center}
April 2005
\end{center}
\vspace*{1.0cm}
\newpage

Quasinormal modes associated with black hole metric perturbations \cite{rg}
were introduced by Vishveshwara \cite{vish} and have been found to be a
useful tool in analyzing the properties of a black hole using an external
probe. The usual calculation of the quasinormal modes in a given geometry
involves finding the solutions of the wave equation with suitable boundary
conditions. In asymptotically flat space-times, the quasinormal modes are
defined as the solutions which are purely ingoing at the horizon and purely
outgoing at infinity \cite{sch}. However, for the case of the BTZ black hole
\cite{btz} which is asymptotically AdS, the potential in the radial part of
the corresponding wave equation diverges at infinity. In this case, the
quasinormal modes are defined as solutions which are purely ingoing at the
horizon and which vanish at infinity \cite{ads,d1}. 

Subsequently, the study of quasinormal modes was given a geometric
formulation following the work of Motl and Neitzke \cite{motl}. The
essential feature of this approach involves the extension of the wave
equation beyond the physical region between the horizon and infinity by
analytically continuing the radial variable $r$ to the whole complex plane. 
For the Schwarzschild case, this leads to a differential equation with
regular singular points at $r = 0,~ 1 {\rm (horizon)~ and}~ \infty$. The
solutions of the wave equation with appropriate boundary conditions are
multivalued around $r = 0$ and $r = 1$, leading to nontrivial monodromies
around suitably chosen closed contours in the complex $r$ plane.  An
equation relating the relevant monodromies then leads to the evaluation of
the quasinormal modes. In the case of the BTZ black hole a similar analysis
involving the monodromies was done by Musiri and Siopsis \cite{mus}. In this
case, the solution of the wave equation vanishes at infinity, which again
leads to an equation relating the monodromies around $r=0$ and $r = 1$.
Solution of this equation again leads to the quasinormal mode frequencies.
In both the approaches mentioned above, appropriate boundary conditions 
are imposed both at the horizon and at infinity.

In a recent development, Birmingham and Carlip \cite{bc} have analyzed the
BTZ black hole perturbations using boundary conditions formulated in terms
of relation between monodromies at the inner and outer horizons. Their
treatment does not refer to any boundary condition at infinity and no such
condition is imposed upon the solutions of the wave equation. These
monodromy conditions however give rise to the usual quasinormal modes of the
BTZ black hole and the standard ADS/CFT correspondence for the BTZ black
hole \cite{bss} holds for the modes derived from these relations
as well.  The application of these monodromy relations to higher dimensional
near-extremal black holes with asymptotically flat geometries, whose
near-horizon region contains the geometry of a BTZ black hole, gives rise to
the so called ``non-quasinormal modes", as the latter are obtained without
imposing any boundary condition at infinity. As emphasized in
\cite{bc}, the conditions on the monodromies in general do not follow from 
the usual boundary conditions of the system, but are somewhat ad hoc.

In this Letter we shall show that the results obtained in Ref. \cite{bc}
follow directly from the properties of the BTZ black hole and the monodromy
conditions assumed there can be justified. The basic feature of the BTZ
black hole which allows this to happen is that the Euclidean BTZ is a
hyperbolic 3-manifold which is geometrically finite \cite{f1}. The three
dimensional hyperbolic structures for such a manifold, according to
Sullivan's theorem \cite{sull}, are in 1-1 correspondence with the two
dimensional conformal structures of its boundary. This is nothing but a
precise mathematical statement of holography for the BTZ black hole
\cite{f1,man}. For the Euclidean BTZ black hole, the two dimensional
conformal structure of its boundary is described by the Teichmuller
parameter $\tau$ of a two torus $T^2$. Two such parameters $\tau$ and
$\tau^{\prime}$ are considered equivalent if they are related by the action
of the modular group. On the other hand, the monodromies of the solutions of
the wave equation at inner and outer horizon are determined by the
hyperbolic structures encoded in the metric of the BTZ manifold. Thus, the
1-1 correspondence between the conformal structures and the hyperbolic
structures following from  Sullivan's theorem requires that the
monodromies be invariant under the action of the modular group as well. We
shall show that imposing this requirement leads to the non-quasinormal
modes of Birmingham and Carlip \cite{bc}. We start with a brief summary of
the properties of the BTZ black hole that is relevant for our purposes.

We shall first review the hyperbolic structure of the BTZ black hole.
The metric of a BTZ black hole with mass $M$ and angular momentum $J$ is 
given by
\be
ds^2 = - \left ( -M + \frac{r^2}{l^2} + \frac{J^2}{4 r^2} \right ) dt^2
       + \left ( -M + \frac{r^2}{l^2} + \frac{J^2}{4 r^2} \right )^{-1} 
       dr^2 + r^2 \left ( d \phi - \frac{J}{2 r^2} dt \right )^2
\ee
where $\Lambda = - {1}{l^2}$ is the cosmological constant and the Newton's 
constant $G$ satisfies the condition $8G = 1$.
The inner and 
outer horizons are given $r= \rho_+$ and $r= \rho_-$ respectively where 
\be
\rho_{\pm} = \left [ \frac{Ml^2}{2} \left ( 1 \pm \sqrt{1 - 
\frac{J^2}{M^2 l^2}} ~ \right ) \right ]^{\frac{1}{2}}.
\ee

	For our purpose it is necessary to consider the Euclidean 
continuation of the BTZ black hole, which is obtained from (1) by setting 
$t = i t_E$ and $J = -i J_E$. The Euclidean metric is then given by
\be
ds_E^2 = \left ( -M + \frac{r^2}{l^2} - \frac{J_E^2}{4 r^2} \right ) 
dt_E^2 + \left ( -M + \frac{r^2}{l^2} - \frac{J_E^2}{4 r^2} \right )^{-1}
       dr^2 + r^2 \left ( d \phi - \frac{J_E}{2 r^2} dt_E \right )^2.
\ee
In the Euclidean case the inner and outer horizons are given by 
\be r_{\pm} = \left [ \frac{Ml^2}{2} \left ( 1 \pm \sqrt{1 + 
\frac{J_E^2}{M^2 l^2}} ~ \right ) \right ]^{\frac{1}{2}}.
\ee
In terms of the variables
\bea
x &=& \left ( \frac{r^2 - r_+^2}{r^2 - r_-^2} \right )^{\frac{1}{2}}
    {\rm cos} \left ( \frac{r_+}{l^2} t_E + \frac{ |r_- |}{l} \phi \right 
) {\rm exp} \left ( \frac{r_+}{l} \phi - \frac{ |r_- |}{l^2} t_E  \right )
\nonumber \\
y &=& \left ( \frac{r^2 - r_+^2}{r^2 - r_-^2} \right )^{\frac{1}{2}}
    {\rm sin} \left ( \frac{r_+}{l^2} t_E + \frac{ |r_- |}{l} \phi \right
) {\rm exp} \left ( \frac{r_+}{l} \phi - \frac{ |r_- |}{l^2} t_E  \right )
\nonumber \\
z &=& \left ( \frac{r^2 - r_+^2}{r^2 - r_-^2} \right )^{\frac{1}{2}}
{\rm exp} \left ( \frac{r_+}{l} \phi - \frac{ |r_- |}{l^2} t_E  \right )
\eea
the Euclidean metric (3) becomes 
\be
ds_E^2 = \frac{l^2}{z^2} (dx^2 + dy^2 + dz^2), ~~ z > 0.
\ee
The metric in (6) is that of the upper half space of the the three 
dimensional hyperbolic space $H^3$ and thus BTZ is locally isometric to 
$H^3$. However, the periodicity of the coordinate $\phi $ leads to the
identifications 
\be
(x,y,z) \sim
e^{\frac{2 \pi r_+}{l}} \left ( x~ {\rm cos} \left ( \frac{2 \pi |r_-|}{l}
\right ) - y~ {\rm sin} \left ( \frac{2 \pi |r_-|}{l} \right ),
x~ {\rm sin} \left ( \frac{2 \pi |r_-|}{l} \right )
+ y~ {\rm cos} \left ( \frac{2 \pi |r_-|}{l} \right ), z \right ).
\ee
In terms of the spherical polar coordinates in the upper half plane given by
\be
(x,y,z) = (R~ {\rm cos} \theta ~ {\rm cos} \xi,~ R~ {\rm sin} \theta ~ {\rm
cos} \xi,~ R~ {\rm sin} \xi ),
\ee
we can write
\be 
ds_E^2 = \frac{l^2}{{\rm sin}^2 \xi}
\left (\frac{dR^2}{R^2} + d \xi^2 + {\rm cos}^ \xi d \theta^2 \right ),
\ee
with the identifications in (7) now being given by 
\be
(R, \theta, \xi) \sim \left ( R {\rm e}^{\frac{2 \pi r_+}{l}}, ~ 
\theta + \frac{2 \pi |r_-|}{l}, ~ \xi \right ).
\ee
The resulting manifold has topology of a solid torus, each constant $\xi
\neq \frac{\pi}{2}$ section of which is a two torus $T^2$ \cite{torus}.

In order to proceed, the next point to note is that the 
BTZ black hole, which is locally isomorphic to $H^3$ 
is geometrically finite \cite{f1}. This feature establishes a geometric 
result that is equivalent to a precise mathematical statement of 
holography. It establishes, by using a theorem 
(Sullivan's theorem) \cite{sull}, the 
equivalence of the hyperbolic structures of the BTZ 3-manifold 
with the conformal structures of its boundary. More precisely, if
$K$ is a geometrically finite hyperbolic  3-manifold with boundary then
 Sullivan's theorem states that as 
long as $K$ admits one hyperbolic realization, there is a 1-1 
correspondence between hyperbolic structures on $K$ and conformal 
structures on its boundary $\partial K$, the latter being the 
Teichmuller space of the boundary $\partial K$.

Let us now analyze the implication of Sullivan's theorem for the BTZ
black hole.
As discussed above, the boundary of the BTZ black hole has the topology of
$T^2$ and the corresponding Teichmuller space is given by a the fundamental 
region of the complex variable $\tau$. In other words, two Teichmuller 
parameters $\tau$ and $\tau^{\prime}$ are equivalent if 
\be
\tau^{\prime} = \frac{a \tau + b}{c \tau + d}, ~~ ad - bc = 1
\ee
and $a,b,c,d \in Z$. The transformation in (11) is nothing but the action 
of the modular group on $\tau$ generated by the operations
\bea
S &:& \tau \rightarrow - \frac{1}{\tau} \nonumber \\
T &:& \tau \rightarrow \tau + 1.
\eea
In terms of the horizon radii of the Euclidean BTZ black hole, 
the effective action of the modular group can be written as 
\bea
S &:& r_+ \leftrightarrow r_- \nonumber \\
T &:& r_+ \rightarrow r_+, ~~ r_- \rightarrow r_- +  r_+.
\eea 
By analytically continuing back to the corresponding Minkowskian 
variables, $S$ and $T$ would have analogous actions on $\rho_{\pm}$ 
respectively. 

According to Sullivan's theorem, the conformal structures at the boundary of
the BTZ black hole are in 1-1 correspondence with the hyperbolic structures
encoded in the BTZ metric. It should thus be possible to have a
corresponding action of the modular group on the quantities determined by
the hyperbolic structure, i.e. the metric of the BTZ black hole. In this
case, such a natural object determined by the metric is given by the
monodromy of the solution of the wave equation in the BTZ background. Below
we briefly recall how such monodromies arise in this context.

Following Birmingham and Carlip \cite{d1,bc}, we consider 
the wave equation for a massless scalar field $\chi$ in the background of 
the BTZ black hole given by
\be
\bigtriangledown^2 \chi = 0.
\ee 
Using the separation of variables
\be
\chi = R (r)~ {\rm e}^{-i w t} ~{\rm e}^{ i m \phi},
\ee
the equation satisfied by the function $R(r)$ can be written in the form of
the standard hypergeometric equation \cite{d1}. Near $r = \rho_+$, the
radial part of the ingoing solution is given by 
\be
R(\rho_+) \sim (r^2 - \rho_+^2)^{- \frac{i}{8 \pi}[\beta_R(w + \frac{m}{l})
~+~ \beta_L(w - \frac{m}{l})]},
\ee
where $\beta_{R,L} = \frac{2 \pi l^2}{(\rho_+ \pm \rho_-)}$.
The monodromy of this solution is given by its change under $2 \pi$ rotation
around $r = \rho_+$ and can be written as 
\be
P(\rho_+) = {\rm exp} \left ( \frac{1}{4} \left [ \beta_R \left (
w + \frac{m}{l} \right ) + \beta_L \left ( w - \frac{m}{l} \right ) 
\right ] \right ).
\ee
The above solution can be analytically continued and near $r = \rho_-$, it
can be expressed as a suitable linear combination of two functions 
\be
R^{\pm}(\rho_-) \sim (r^2 - \rho_-^2)^{\pm \frac{1}{8 \pi}[\beta_R(w +
\frac{m}{l}) ~-~ \beta_L(w - \frac{m}{l})]}.
\ee
The monodromies of these solutions at $r = \rho_-$
are given by
\be
P^{\pm} (\rho_-) =
{\rm exp} \left (\pm \frac{1}{4} \left [ \beta_R \left (
w + \frac{m}{l} \right ) - \beta_L \left ( w - \frac{m}{l} \right )
\right ] \right ).
\ee

At this stage, Birmingham and Carlip \cite{bc} imposed the following
conditions on the monodromies at the inner and outer horizons:
\be
P (\rho_+)~P (\rho_-) = 1.
\ee
It was also shown by them that this condition is satisfied if either
\be
P (\rho_+)~P^+ (\rho_-) = 1
\ee
or
\be
P (\rho_+)~P^- (\rho_-) = 1.
\ee
These conditions on monodromies are clearly different from the usual
Dirichlet boundary condition imposed at infinity. Due to this reason the 
frequencies obtained from the above equations are called the non-quasinormal
mode frequencies and are given by \cite{bc}
\be
w_L = \frac{m_L}{l} - \frac{4 \pi i}{\beta_L} n
\ee
and
\be
w_R = \frac{m_R}{l} - \frac{4 \pi i}{\beta_R} n
\ee
where $n \in Z$ and $m_L,~m_R \in Z$. It may be noted that the AdS/CFT
correspondence discussed in ref. \cite{bss} is consistent with these 
non-quasinormal mode frequencies as well.

Our goal now is to provide a justification for the eqns. (20) - (22) using
the geometric finiteness property of the BTZ black hole.  As argued before, 
Sullivan's theorem in this case leads to a natural action of the modular
group on the monodromies. Since the Teichmuller parameters related by the
action of the modular group are considered equivalent, Sullivan's
theorem therefore suggests that the monodromies should be left invariant
under the action of the modular group.

Let us now analyze the implication of the above assertion. Consider the
action of the generator $S$ of the modular group on the monodromy $P
(\rho_+)$. As a result of our assertion, we demand that 
\be
S P (\rho_+) = P (\rho_+), 
\ee
i.e. the monodromy is kept invariant under the action of the modular group. 
However, using (13) we get
\be
S P (\rho_+) = P (S \rho_+) = P (\rho_-)
\ee
However, for $P (\rho_-)$, we could take either $P^+ (\rho_-)$ or 
$P^- (\rho_-)$. Therefore, the invariance of the monodromies under the
action of the modular group leads to the equations 
\be
P (\rho_+) = P^+ (\rho_-)
\ee
or
\be
P (\rho_+) = P^- (\rho_-),
\ee
which are the same as imposed in ref. \cite{bc} and they lead to the same 
non-quasinormal mode frequencies as given in (23) and (24). Using the above 
values of $w_L$ and $w_R$ it is also easy to see that the action of $T$ on the 
monodromies do not lead to any new condition.
  
We have therefore shown that the conditions on the monodromies used to
evaluate the frequencies of the non-quasinormal modes 
introduced in ref. \cite{bc} follow from the geometric properties of the
BTZ black hole. Our analysis is based on the observation made in \cite{f1} 
that BTZ black hole
is geometrically finite and thus Sullivan's theorem is applicable in
this case. Using Sullivan's theorem, the hyperbolic structures of the BTZ
black hole are related to Teichmuller parameters of its boundary, which is 
$T^2$. On the Teichmuller space of $T^2$, there is a natural action of
the modular group and two different Teichmuller parameters related by the
action of the modular group are equivalent. As a consequence of Sullivan's
theorem, this leads to the condition that the monodromies be invariant under
the action of the modular group. This invariance condition on the
monodromies immediately leads to the relations between the monodromies that
were imposed in ref. \cite{bc} in an ad hoc fashion. We are thus able to
give a geometric interpretation of the monodromy conditions imposed by
Birmingham and Carlip. Moreover, this geometric condition is strongly
related to the holographic nature of the BTZ black hole. It may also be
noted that a variety of black holes arising from string theory have a
near-horizon geometry containing the BTZ black hole. It is thus plausible
that our analysis applies to this large class of black holes as well.


\vskip 1 cm

\noindent
{\bf Acknowledgments} : A part of this work was done during KSG's visit to
the Hamilton Mathematics Institute, Trinity College, Dublin, Ireland and
Abdus Salam International Centre for Theoretical Physics, Trieste, Italy.
KSG would like to thank the Hamilton Mathematics Institute and the
Associateship Scheme of the Abdus Salam ICTP for financial support.

\bibliographystyle{unsrt}

\end{document}